\RequirePackage{lineno} 
\documentclass[%
letterpaper,
reprint,
superscriptaddress,
showpacs,
showkeys,
amsfonts,amsmath,amssymb,
aps,
prb,
floatfix
]{revtex4-1}
\usepackage{graphicx}
\usepackage{dcolumn}
\usepackage{bm}

\newcommand{\ba}{\begin{align}}

\begin{document}
\title{Band alignment and directional stability in abrupt and polar-compensated Si/ZnS interface calculations}

\author{D.~H.~Foster}
\affiliation{Department of Physics, Oregon State University, Corvallis, Oregon 97331, USA}

\author{G.~Schneider}
\email{Guenter.Schneider@physics.oregonstate.edu}
\affiliation{Department of Physics, Oregon State University, Corvallis, Oregon 97331, USA}

\date{\today}

\begin{abstract}
%
%
We perform a first principles investigation of Si/ZnS interface properties for the [111], [100], and [110] directions, including single-substitution polar-compensated interfaces.
The asymmetry of general interface directions within semiconductor or oxide componds
poses known challenges for standard methods of calculation:
a multiplicity of interface distinctions, artificial electric fields, and indeterminacy of orientation stability.
By placing each distinct interface in a variety of supercell environments, we demonstrate that the spread of both band offsets and interface enthalpies is acceptably small for reasonable cell lengths, removing the need for corrections involving inappropriate assumptions or computationally expensive structures.
Both the orientation and the ionic character of abrupt (111) zinc blende interfaces are shown to affect band alignment and interface enthalpy.
%
%
%
We find that the band offsets for the compensated and abrupt (111) and (100) interfaces lie on a strongly bimodal distribution of total width greater than 1.2 eV, while the (110) band offset lies near the distribution midpoint.
The midpoint agrees with previous experiments on (100) interfaces, but only one peak of the distribution agrees with (111) interface experiments, indicating that the grown macroscopic (111) interfaces had significant selectivity among the possible microscopic interfaces.
The polar-compensated interfaces are shown to be more stable than the corresponding abrupt interfaces over most growth conditions.
%
\end{abstract}




\maketitle

\section{\label{s:intro}Introduction}

Surfaces, interfaces, and nanostructures of complex metal oxides are of current interest\cite{goniakowski08pol,chambers10ins,moetakef11ele,huang12map} in part due to the interplay of ionic and electronic reconstruction that occurs at crystal terminations of almost all orientations.
However, polar or nominally polar interfaces between semiconductors contain many of the features of their oxide counterparts, and have been studied\cite{baraff77sel,harrison78pol,kley94ato,bernardini98mac,leitsmann06str} for several decades.
In both cases, ideal abrupt polar interfaces having only electronic reconstruction are found to be unstable with very few exceptions.
Typically there exist lower energy (enthalpy)
compositional reconstructions which reduce or remove the termination-dependent compositional bound charge\cite{stengel11fir} that renders an interface ``polar''.
Computationally\cite{chambers10ins} this has been the case even for the well known interface\cite{ohtomo04ahi} between perovskites SrTiO$_3$ (STO) and LaAlO$_3$ (LAO) that exhibits free electronic interface charge.

ZnS is lattice matched to Si within $0.4\%$ and can be epitaxially grown by molecular beam epitaxy\cite{maierhofer91val,romano95int,zhou97epi} (MBE), pulsed laser deposition\cite{shen94cry}, and chemical vapor deposition\cite{wei02dep,*tran02isl,*lee03str} in directions in which abrupt IV/II-VI interfaces are strongly polar.
Applications of this heterojunction in electronic\cite{yu13lar}, optoelectronic\cite{mastio01lat,xu09str}, and photovoltaic\cite{hsiao13pre} devices depend on the screened interface charge, the interface electrostatic dipole, and any contributions\cite{wang12exc}, in addition to the dipole, to the difference in the local vacuum potentials (LVP) of the two materials.
The equilibrium interface charge goes to zero in sufficiently large devices due to screening, while the LVP offset is typically nonzero, even at non-polar interfaces.
The LVP offsets  are highly dependent on the interface structure, which in turn depends on which configurations have the lowest energies.
Studies of energy-driven structural properties are are often weak in specific parameter prediction, because only a small set of possible defect structures can be considered computationally.
Nevertheless, clustering of energies and offsets within a modest data set can be valuable for understanding the macroscopic interface.

Even if restricted to abrupt interfaces normal to a given unsigned direction, band offsets are non-transitive
and generally depend on the orientation sign and terminating ion type of the interfaces.
The lack of a generalized mirror symmetry along a \emph{general} direction for a zinc blende (ZB) crystal
has a number of interesting consequences: (a) there are four terminations (two dangling bond configurations times two terminal ion choices),
 (b) it poses challenges for band offset calculations due to nonzero bulk electric fields $\mathcal{E}$ in two-slab supercells, and (c) it imposes rigid algebraic restrictions which prohibit the calculation of absolute interface energies using quasi-1D supercell computations\cite{rapcewicz98con,chetty92gas,bylander88lar}.
In many interface studies\cite{dandrea90sta,kley94ato}, band offset calculations for directions lacking mirror symmetry are not attempted.
One can eliminate artificial electric fields at the expense of adding two vacuum interfaces, with or without\cite{mankefors99sch} adding a dipole correction to the potential.
Alternatively, Bernardini and Fiorentini\cite{bernardini98mac} have developed a method which spatially folds the supercell charge density to obtain a single composite interface dipole magnitude for the two interfaces.
It appears that existing methods\cite{bernardini98mac,leitsmann06str} designed to handle $\mathcal{E} \neq 0$ with two-slab supercells effectively use a transitivity approximation, assuming the LVP offsets at opposing interfaces to have a single value.
The energy determination problem requires either a large scale 3D supercell calculation or a local energy density approach such as the Voronoi polyhedra technique developed by Rapcewicz \emph{et al}.\cite{rapcewicz98con}
Leitsmann \emph{et al}.\cite{leitsmann06str,leitsmann07ele} compare several approaches to energy and band offset calculations.

For Si/ZnS interfaces, band offset measurements have been conducted for the [111]\cite{maierhofer91val} and [100]\cite{brar98ban} directions, while the [110] band offset has been calculated\cite{li96ban}.
Maierhofer \emph{et al}.\cite{maierhofer91val} have additionally provided valuable insight into the structure of MBE-grown (111) interfaces.

In this work we use calculations based on density functional theory to examine interfaces between Si and ZnS having the polar directions $[111]$ and $[100]$, and the non-polar direction $[110]$.
In the approach we take, all distinct interfaces are treated individually with respect to both valence band offsets (VBO or $\Delta_V$) and energies $E$.
The dependence of interface properties on signed orientation and chemical termination plays a central role in our findings.
Using supercell total energy calculations, we determine absolute or relative energies for abrupt interfaces, as well as for several polar-compensated interfaces having a single atomic substitution in a minimal surface unit.
The set of supercells overdetermines the calculable energies, allowing us to compare different values resulting from selected subsets of supercells.
Energies are determined for Zn-rich and Zn-poor growth conditions.
We estimate band offsets using an extrapolation of bulk reference potentials, which vary linearly due to electric fields.
Multiple estimates for the band offsets of each interface are determined using various supercells having different opposing interfaces and electric field magnitudes.
We also use vacuum regions with pseudo-hydrogen passivated surfaces to provide additional checks for the calculated offset values.
A sufficiently large supercell length $L$ is chosen so that definite conclusions may be drawn without the need of numerous finite size corrections.

The well studied oxide interfaces having free charge
have a theoretical decomposition of charge components that is applicable to general semiconductor and oxide interfaces.
The total interface charge $\sigma$ has contributions from free charge $\sigma_{\textrm{free}}$, induced bound charge $\sigma_{\chi}$, and non-induced bound charge (compositional charge $\sigma_0$ plus any piezoelectric, pyroelectric, or ferroelectric charges).\cite{stengel11fir}
One may decompose the total charge into these components and relate the charges to the interface theorem\cite{vanderbilt93ele} from the modern theory of polarization.\cite{king-smith93the} We leave this investigation, as well as discussions of the superlattice band structures, to future work. We note that at least some polar superlattices, although having metallicity and band hybridization between nominally unoccupied surface states and valence band states, allow the interface free charge to be unequivocally determined directly from the occupation of Kohn-Sham states $\psi_{n\bm{k}}$ having surface state character (unpublished).\footnote{For the abrupt stoichiometric [111] superlattice having interface bonds along [111], nearly all non-empty calculated states $\psi_{n\bm{k}}$ have either very strong localization or negligible amplitude at the donor interface.
The apparent thinness of the shells of $k$-space in which hybridization might be observable may be due to the fact that the apparent crossings of the donor interface state band all lie relatively close to $\Gamma$ (within 30\% of the 2D Brillouin zone effective radius) where the supercell Hamiltonian has significant symmetry $(C_{3v})$.
The lack of hybridization is noted at $k$-point densities as high as $36\times36\times1$ and is aided by using the tetrahedral method of $k$-point smearing.
}

The remainder of this Article is structured as follows.
Section \ref{s:studies} discusses previous work on Si/ZnS interfaces.
Sections \ref{s:interfaces} and \ref{s:superlattices} describe abrupt and polar compensated zinc blende interfaces, and the relation of band offsets, charge transfer, and electric fields in superlattices.
Section \ref{s:energiestheory} describes the calculation of chemical potential bounds and interface energies.
Section \ref{s:offsetdetermination} describes our method for band offset determination.
Details of the calculation methods are given in Sec.~\ref{s:methoddetails}.
Sections \ref{s:ifcenergyresults} and \ref{s:bandoffsetresults} describe the results for interface energies and band offsets, respectively.
Section \ref{s:conclusions} gives our conclusions.

\section{\label{s:theory}Background and Methods}

\subsection{\label{s:studies}Previous Studies of Si/ZnS Interfaces}

Ultraviolet and X-ray photoemission measurements by Maierhofer \emph{et al}.\cite{maierhofer91val} on ZnS grown by MBE on cleaved (111) (2$\times$1) Si surfaces yield a Si-to-ZnS VBO value of $\Delta_V = -0.7 \pm 0.2$ eV.
The work mentions a possible alternative methodological interpretation which could the lower the mean value, possibly to $-1.0$ eV.
Low energy electron diffraction (LEED) measurements reported in the work show that no regular surface structure larger than the (1$\times$1) primitive cell exists upon sub-monolayer deposition.
Photoemission results suggest the sub-monolayer contains more S than Zn, which is corroborated by studies for (100) growth of ZnS\cite{zhou97epi} and ZnSe\cite{bringans89bon} noting an anion sticking preference and a Si-anion chemical reaction.
For the growth of ZnS onto the (7$\times$7) reconstructed Si $(111)$ surface, which is less smooth than the (2$\times$1) reconstructed surface, Maierhofer \emph{et al}.~observed indirect evidence of a silicon sulfide layer (nominally SiS$_2$).
The evidence of penetrating S atoms comes from deformation of the Si-$2p$ core level photoemission peak in a manner that is consistent with the presence of high oxidation states for Si.

Electrical measurements\cite{brar98ban} of a MBE-grown $(100)$ interface indicate $\Delta_V = -1.4$ eV.
Photoemission evidence of a SiS$_2$ layer has been observed\cite{weser87pho} for deposition of S on Si $(100)$ at elevated substrate temperatures (200 $^{\circ}$C) but not at room temperature.
Zhou \emph{et al}.\cite{zhou97epi}~measured more S than Zn adsorbing onto a Si (100) (2$\times$1) surface at 340 $^{\circ}$C during the sub-monolayer coverage period.
Evidence of Si-S chemical reactivity was present, and was reduced when ZnS was deposited onto an As-passivated Si surface.

Theoretical treatments\cite{li96ban,jedrzejek95for,lewyanvoon97ele,wang96sup,*laref06ele,*rhim09int} of Si/ZnS and Si/ZnSe interfaces and superlattices are most often focused on superlattice electronic structure and/or thin superlattices.
Li \emph{et al}.\cite{li96ban} calculate $\Delta_V = -1.5$ eV for (110) Si/ZnS, while first principles offset calculations in other directions are unknown to the authors.

\subsection{\label{s:interfaces}Interface Terminations and Polar-Compensated Interfaces}

Abrupt diamond/ZB interfaces can be specified by an unsigned direction,
and the specific termination of the ZB compound.
To specify terminations uniquely for the [111] and [100] directions, we use the notation $c_b$ ($a_b$) to specify a cation (anion) termination with $b=1,2,3$ dangling bonds. The $b=1$ dangling bond lies along the [111] direction.
The $c_{1}$ and $a_{3}$ terminations of zinc blende correspond to the interface orientation commonly denoted $[111]A$, and the $a_{1}$ and $c_{3}$ terminations correspond to orientation $[111]B$.
Both $b=2$ terminations correspond to the [100] direction.
The lone (110) termination is the only non-polar abrupt interface that we discuss.

\begin{figure}
\includegraphics{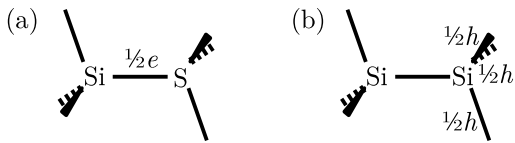}
\caption{\label{f:Si_S_compensation}(a) The $a_{1}$ (111) Si/ZnS interface, showing the extra half electron that has no low energy bonding state, and may partially leave the interface vicinity, leaving a positive charge. (b) The Si$_{\textrm{S}}$ substitution, which creates $3/2$ holes. The $n_a$ interface has one Si$_{\textrm{S}}$ substitution for every four Si-S bond sites and allows all extra electrons to fill nearby holes.}
\end{figure}

We consider the most primitive set of ion-substitution polar-compensated interfaces\cite{harrison78pol} for the [100] and [111] directions.
Figure \ref{f:Si_S_compensation} shows the (111) $1/4$ Si$_{\textrm{S}}$ substitution, a (2$\times$2) polar-compensating substitution of the $a_{1}$ interface which alters $1/4$ of the atoms in the first plane on the ZnS side of the interface.
Other (111) compensations we examine are the $1/4$ Zn$_{\textrm{Si}}$ substitution for the $a_{1}$ interface and the $1/4$ Si$_{\textrm{Zn}}$ and $1/4$ S$_{\textrm{Si}}$ substitutions for the $c_{1}$ interface.
The four $1/4$ substitutions mentioned above can be respectively denoted $n_a$, $c_a$, $n_c$, and $a_n$, with $n$ denoting the ``neutral'' Si.
For the [100] direction we examine the two polar-compensating $1/2$ substitution interfaces having a ($\sqrt{2}$$\times$$\sqrt{2}$) interface unit.
One interface can be considered either a $n_c$ substitution or a $c_n$ substitution.
The other is $n_a$ or $a_n$.

\subsection{\label{s:superlattices}Superlattice Potential and Charge}

The periodic boundary conditions within a supercell or superlattice are well known to cause artificial bulk electric fields.
The piecewise-linear approximation (\emph{cf}.~Ref.~\onlinecite{kley94ato}) of a superlattice potential profile obeys
\footnote{
Some degree of approximation, even for $L\rightarrow \infty$, is likely involved in Eq.~(\ref{e:supercellperiodicity}).
Replacing the LVP offsets in Eq.~(\ref{e:supercellperiodicity}) with the negatives of the electrostatic potential offsets results in a strictly valid equation for $L\rightarrow \infty$.
Unlike the electrostatic potential offsets, the LVP offsets $\Delta$ include effects from exchange-correlation\cite{wang12exc} and allow charge relaxation effects.
}
 the $z$-periodicity relation
\ba
\mathcal{E}_1 l_1 + \Delta_{12,1} + \mathcal{E}_2 l_2 - \Delta_{12,2} = 0, \label{e:supercellperiodicity}
\end{align}
along with the the electrostatic relation
\ba
\mathcal{E}_2 - \mathcal{E}_1 = 4 \pi \sigma_{1} = -4 \pi \sigma_{2}. \label{e:sigmatotal}
\end{align}
Here $\mathcal{E}_j$ is the bulk electric field in slab $j$, the interface $j$ lies on the side of slab $j$ in the positive, or right, direction, and $\Delta_{kl,j}$ is the jump in the electron LVP going from slab $k$ to slab $l$ at interface $j$.
$l_j$ is the length of slab $j$.
$\sigma_{j}$ is the total (screened) interface charge density at interface $j$.
The offsets $\Delta$ are considered to be independent of bulk electric fields, and are related to the VBO and conduction band offset (CBO) through the bulk electron affinities and bulk band gaps of the two materials.
We typically take the $z$ coordinate to be in a direction normal to the interfaces under consideration.

The valence band at one interface cannot be higher than the conduction band at the other interface (except perhaps as allowed by finite size effects).
With or without the presence of interface states, which enter the gap and have partial occupations, the potential height change $|\mathcal{E}_j| l_j$ is bounded so that $|\sigma| \sim L^{-1}$. $|\sigma|$ can often be interpreted as a net charge transfer from a donor interface to an acceptor interface.

\subsection{\label{s:energiestheory}Interface Energies}
The sum of the two interface energies $E_{I1}$ and $E_{I2}$ in a two-slab supercell is estimated as
\ba
E_{I1} + E_{I2} = E_{\Omega} - \sum_{\alpha} n_{\alpha} \mu_{\alpha}, \label{e:sumifcenergies}
\end{align} 
where $E_{\Omega}$ is the calculated supercell energy and $n_{\alpha}$ is the number of atoms (or small units such as Zn$_1$S$_1$) of type $\alpha$ comprising the cell. $\mu_{\alpha}$ is the chemical potential of the constituent $\alpha$.
The existence of bulk Si and bulk ZnS provide the constraints $\mu_{\textrm{Si}} = E_{b,\Omega}(\textrm{Si})$, $\mu_{\textrm{ZnS}} =  E_{b,\Omega}(\textrm{ZnS})$,
where $E_{b,\Omega}(X)$ is the calculated bulk energy of the material unit $X$, constrained within the cross-section of the supercell.
The Zn-rich and Zn-poor conditions are 
\ba
\mu_{\textrm{Zn}}\big|_{\textrm{Zn-rich}} &= E_b(\textrm{Zn}), \label{e:muZn_M}\\
\mu_{\textrm{S}}\big|_{\textrm{Zn-poor}} &= (E_b(\textrm{SiS}_2) - \mu_{\textrm{Si}}) / 2. \label{e:muS_barM}
\end{align}
$E_b$ denotes a calculated zero strain bulk energy; strained bulk energies for Zn and SiS$_2$ are not calculated because the constrained structures are not known.
Regardless of growth conditions,
\ba
\mu_{\textrm{ZnS}} = \mu_{\textrm{Zn}} + \mu_{\textrm{S}}. \label{e:muZnSsum}
\end{align}
Stoichiometric supercells have 
$n_{\textrm{Zn}} = n_{\textrm{S}}$
and thus do not depend on the growth condition coordinate,
\ba
\mu_x \equiv \mu_{\textrm{Zn}} - \mu_{\textrm{S}}.
\end{align}

The lack of an intrinsic (bulk) orientation in ZB along the [110] and [100] directions allows two-slab supercells to be constructed so that the two interfaces are mirror images, up to operations in the transverse plane.
This generalized mirror symmetry implies $E_{I1} = E_{I2}$, 
and permits absolute energy calculations.
Zinc blende interfaces in general directions, which in this sense includes [111], do not allow a such a construction.\cite{chetty92gas,bylander88lar} 
Individual ZB/ZB and ZB/diamond (111) interface energies do have well defined values\footnote{Wurtzite and other structures belonging to 10 particular point groups, have directions in which they do not have unique surface or interface energies.\cite{rapcewicz98con}}, but they cannot be determined by total energy calculations using an set of quasi-1D supercell total energies, even allowing for vacuum slabs and slabs of auxiliary materials.
As mentioned in Sec.~\ref{s:intro}, there are two methods of calculation: (a) a full 3D calculation of an embedded ZB crystal, such as a tetrahedral crystal with faces (111), (100), (010), and (001), and (b) local energy density methods.
In the present work we must be content with interface energies relative to others having the same interface orientation.

\subsection{\label{s:offsetdetermination}Band Offsets}

Typical first principles VBO calculations\cite{vandewalle87the} use the supercell-calculated slab potentials $V_1$ and $V_2$, and the bulk-calculated quantities $E_{V1}-V_1$ and $E_{V2}-V_2$, where $E_{Vj}$ is the valence band maximum (VBM) level for slab $j$.
The CBO and the LVP offset $\Delta$ are then easily obtained using experimental band gaps and electron affinities (for these we use values from Ref.~\onlinecite{maierhofer91val}).
For ZB (111) interfaces, one encounters the problem that the $V_j$ are not constant in the bulk regions, due to nonzero electric fields $\mathcal{E}_j$.
In this case we take the very simple approach of defining nominal interface positions $z_{Ij}$ for the relaxed structure, and determine the reference potential offset at interface $I_j$ by linear extrapolation of the potentials. For interface $I_1$,
\begin{align}
\Delta_{V,1} &\equiv V_2(z_{I1}) - V_1(z_{I1}) \nonumber \\
&= [V_2(z_{c2}) + \mathcal{E}_2(z_{c2}) (z_{I1}-z_{c2})] - \nonumber \\*
&\qquad [V_1(z_{c1}) + \mathcal{E}_1(z_{c1}) (z_{I1}-z_{c1})], \label{e:potlinextrap}
\end{align}
with a similar expression holding for $I_2$.
Here $z_{cj}$ is a position near the center of slab $j$.
The potential may be examined using a single or double\cite{baroni89can} convolution smoothing method.
We take the nominal interface positions to be midway between the atomic planes  that form the interface.
(A plane's $z$ position may involve averaging over relaxed ion positions.)
In our calculations $V_j(z)$ is the negative of the physical electrostatic potential due to valence charge density and ion core charges.

The extrapolation produces an $O(L^{-1})$ uncertainty in $\Delta_{V,1}$ of approximately
\ba
|\mathcal{E}_1 - \mathcal{E}_2| \delta(z_{I1}) / 2. \nonumber
\end{align}
$\delta(z_{I1})$ is an estimated width based perhaps on the calculated set of supercell reference potentials $V(z)$.
We do not attempt to determine this uncertainty directly, but rely on the \emph{a posteriori} spread of offset values, having typically four offset calculations per interface.
The spread of $\Delta_V$ for a single interface, typically 0.1 to 0.3 eV, is small relative to the range of $\Delta_V$ for different interfaces corresponding to a single direction ($\sim 1$ eV).
Thus, despite the dipole moment being formally ill-defined for a charged region of space, the extrapolation determined dipoles of well-defined (very small charge) and ill-defined distributions agree rather well.



\subsection{\label{s:methoddetails}Methods of Calculation}
The GGA+$U$ method used in our calculations uses the Perdew-Burke-Enzerhof (PBE) GGA functional\cite{perdew96gen} with a value of $U=6$ eV assigned to the Zn-$d$ orbitals\cite{dudarev98ele}.
This value of $U$ both improves the positions of the Zn-$d$  bands and the calculated heat of formation of binary Zn compounds\cite{lany08sem}.
We use the projector augmented wave (PAW) method\cite{bloechl94pro} as implemented in the code VASP\cite{kresse96eff2,*kresse99fro}, with the tetrahedral $k$-point occupation method\cite{blochl94imp}.
The bulk permittivity $\epsilon$ is calculated using density functional perturbation theory.\cite{gonze97dyn}

\begin{figure}
\includegraphics{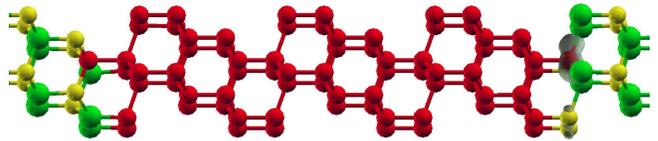}
\caption{\label{f:SionSstate}(Color online) The [111] $(n_c,n_a)$ supercell (Si: red, Zn: green, S: yellow). The charge density (transparent gray) of the highest valence band (all $k$-points) is localized on the three compensating Si-Zn bonds at the $n_a$ interface. This superlattice is fully insulating due to the Si substitutions at both interfaces.}
\end{figure}

The cross sections of all cells and supercells are determined from the GGA+$U$-relaxed Si structure (calculated lattice parameter $a_{\textrm{Si}} = 5.469$ \AA).
For each supercell direction, the corresponding ZnS unit cell is relaxed in the longitudinal direction $z$.
The calculated strain in the transverse directions is $1.012$ ($a_{\textrm{ZnS}} = 5.404$  \AA).
The resulting cross sectional area $S$ of the (1$\times$1) surface units are $21.15$ \AA$^2$, $14.95$ \AA$^2$, and $12.95$ \AA$^2$ for the [110], [100], and [111] directions, respectively.
Supercell length can be described by the number $N$ of 1D quasi-unit cells in each slab, where a quasi-unit cell is the smallest period for which plane-averaged quantities such as $V(z)$ and $dV(z)/dz$ are repeated in $z$.
Supercells\footnote{(1$\times$1) unit cells in both the [110] and [100] directions have two quasi-unit cells and four atoms.
We use hexagonal (1$\times$1) [111] unit cells having three quasi-unit cells and six atoms.} having only abrupt interfaces have (1$\times$1) cross sections while
[100] or [111] supercells having at least one defect interface have ($\sqrt{2}$$\times$$\sqrt{2}$) or (2$\times$2) cross sections, respectively.
An example of a supercell having polar compensation is shown in Figure~\ref{f:SionSstate}.
The transverse linear $k$-point density is chosen so that (1$\times$1) [111] and (1$\times$1) [100] cells have a 12$\times$12 $k$-point grid.

We relax all ion positions in each supercell calculation.
Fully breaking ion symmetries and/or relaxing supercell length are not done for most calculations, as the effects on energies and offsets are less than 0.1 eV and 0.05 eV, respectively.


The inclusion of the primitive single-substitution defects will allow us to eliminate the abrupt interfaces as stable solutions under most growth conditions.
Another benefit of including these polar-compensating defects is the increased number and variety of supercells available for calculation.
The resulting spread of electric fields and interface charges aids both offset and energy calculations.
Four different \emph{systems} of 
equations, each equation (of form Eq.~(\ref{e:sumifcenergies})) representing a [100] supercell calculation having $N\approx16$, are used to produce four independently calculated absolute energy values for each (100) interface.
The same is done with $N=9$ [111] supercells to calculate four independent energy values, relative to $E_{a1}$ or $E_{c1}$,  for each (111) interface.
Band offsets are calculated using Eq.~(\ref{e:potlinextrap}) for each interface in each supercell calculation.

\section{\label{s:results}Results and Discussion}

\subsection{\label{s:ifcenergyresults}Interface Energy Results}


The sum of interface energies in each supercell is given in Table \ref{t:ifcdata}.
The calculated interface energy for the (110) abrupt interface is 2.81 eV/nm$^2$, using $N=20$.

\begin{table}
\begin{ruledtabular}
\begin{tabular}{lrD{.}{.}{2}D{.}{.}{2}D{.}{.}{2}D{.}{.}{2}D{.}{.}{2}D{.}{.}{1}D{.}{.}{1}}
$I_1$, $I_2$ & $N$ & \multicolumn{1}{c}{$\Delta_{V,1}$} & \multicolumn{1}{c}{$\Delta_{V,2}$} & \multicolumn{1}{c}{$\mathcal{E}_1$} & \multicolumn{1}{c}{$\mathcal{E}_2$} & \multicolumn{1}{c}{$\sigma_1$} & \multicolumn{1}{c}{$\Sigma E_{Ij}^{-}$} & \multicolumn{1}{c}{$\Sigma E_{Ij}^{+}$} \\
\hline
[110] & & & & & & & & \\
 & 20 & \multicolumn{2}{c}{$-1.56$} &  \multicolumn{2}{c}{0} &  0 & \multicolumn{2}{c}{5.61} \\
\hline
[100] & & & & & & & & \\
$a_2$, $c_2$ & 16 & -2.11 & -1.11 & -2.66 &  5.09 &  4.29 & 15.4 & 15.4 \\
$a_2$, $a_2$ & 16 & \multicolumn{2}{c}{2.12} & \multicolumn{2}{c}{0} & 0 & 17.2& 22.8 \\
$a_2$, $n_c$ & 16 & -2.13 & -2.16 & -0.36 &  0.32 &  0.38 & 14.0 & 16.8 \\
$a_2$, $n_a$ & 16 & -2.38 & -0.97 &  0.30 &  2.97 &  1.47 & 14.3 & 17.1 \\
$c_2$, $c_2$ & 16 & \multicolumn{2}{c}{$-0.73$} & \multicolumn{2}{c}{0} & 0 & 14.5 & 8.9 \\
$c_2$, $n_a$ & 16 & -0.73 & -0.96 &  1.46 & -0.86 & -1.28 & 13.1 & 10.3 \\
$n_c$, $n_c$ & 16 & \multicolumn{2}{c}{$-2.19$} & \multicolumn{2}{c}{0} & 0 & 11.0 & 11.0 \\
$n_c$, $n_a$ & 16 & -2.16 & -0.76 &  1.06 &  2.17 &  0.61 & 11.3 & 11.3 \\
$n_a$, $n_a$ & 16 & \multicolumn{2}{c}{$-0.74$} & \multicolumn{2}{c}{0} & 0 & 11.8 & 11.8 \\
\hline
[111] & & & & & & & & \\
$a_1$, $c_1$ &  {\em 9} & -2.14 & -0.82 & -4.50 &  9.29 &  7.62 & 10.6 &  10.6\\
$a_1$, $c_1$ & {\bf 12} & -2.19 & -0.70 & -2.60 &  6.67 &  5.13 & 10.7 & 10.7\\
$a_1$, $c_1$ & 18 & -2.07 & -0.69 & -1.26 &  3.74 &  2.76 & 10.9 & 10.9 \\
$a_1$, $c_1$ & 24 & -2.17 & -0.61 &  0.73 &  2.83 &  1.96 & 11.0 & 11.0  \\
$a_1$, $a_3$ & {\em 9}  & -2.19 & -1.92 &  0.34 &  0.62 &  0.16 & 18.7 & 25.1 \\
$a_1$, $a_3$ & {\bf 12} & -2.20 & -1.82 &  0.15 &  0.84 &  0.38 & 18.7 & 25.1 \\
$a_1$, $a_n$ &  9 & -2.19 & -0.90 & -2.75 &  7.39 &  5.60 & 10.0 & 11.6 \\ 
$a_1$, $n_c$ &  9 & -2.09 & -2.03 & -0.99 &  1.23 &  1.23 & 9.8 & 11.4 \\
$a_1$, VL    & 12 & -2.16 &       &  0.05 &  0.99 &  0.52 & & \\
$a_3$, $c_3$ & {\em 9} & -1.62 & -1.42 & -3.40 & 4.63 & 4.44 & 24.8 & 24.8 \\
$a_3$, $c_3$ & {\bf 12} & -1.94  & -1.29 & -2.36 & 4.17 & 3.61  & 24.9 & 24.9 \\
$a_3$, $c_n$ &  9 & -1.74 & -2.12 & -0.62 & -0.72 & -0.06 & 17.8 & 22.7 \\
$a_3$, $n_a$ &  9 & -1.90 & -1.12 & -7.84 &  3.63 &  2.44 & 18.1 & 22.9 \\
$c_1$, $c_3$ & {\em 9} & -0.79 & -1.22 & 0.67 & -2.05 & -1.50 & 17.7 & 11.3 \\
$c_1$, $c_3$ & {\bf 12}& -0.69 & -1.24 & 0.04 & -1.45 & -0.82 & 17.7 & 11.3 \\
$c_1$, $c_n$ &  9 & -0.78 & -2.04 &  1.92 & -6.34 & -4.56 & 10.6 & 9.0 \\
$c_1$, $n_a$ &  9 & -0.87 & -0.91 &  2.65 & -2.90 & -3.06 & 11.0 & 9.4 \\
$c_1$, VL    & 12 & -0.76 &       & -0.05 &  0.41 & -0.25 & & \\
$c_3$, $a_n$ & 12 & -1.27 & -0.74 &  0.28 &  1.12 &  0.46 & 16.9 & 12.1 \\
$c_3$, $n_c$ & 12 & -1.23 & -2.10 & -0.68 & -1.60 & -0.51 & 16.5 & 11.6 \\
$c_n$, $a_n$ &  9 & -2.16 & -0.83 & -0.02 & 5.05 &  2.89  & 9.8 & 9.8 \\
$c_n$, $n_c$ &  9 & -2.09 & -2.07 & -0.15 & 0.23 &  0.21 & 9.5 & 9.5 \\
$n_a$, $a_n$ &  9 & -0.93 & -0.84 & -1.47 & 1.85 & 1.83 & 10.1 & 10.1 \\
$n_c$, $n_a$ &  9 & -2.16 & -0.89 & 1.79 & 2.73 & 0.52 & 9.6 & 9.6
\end{tabular}
\end{ruledtabular}
\caption{\label{t:ifcdata}Valence band offsets and supercell properties.
$\Sigma E_{Ij}^{-}$ and $\Sigma E_{Ij}^{+}$ are the totals of the two supercell interface energies under Zn-poor and Zn-rich conditions, respectively.
VL denotes a vacuum layer.
Units are as follows: valence band offsets $\Delta_{V,j}$ are in eV, electric fields $\mathcal{E}_j$ are in ($10^{-2}$ eV/\AA), and $I_1$ $(I_2)$ total charge density $\sigma_1$ ($-\sigma_1$) is in ($10^{-4}$ e/\AA$^2$). Interface energy quantities are in (eV/nm$^2$).
Among supercells having the same two interface types, data from only one calculation, denoted by a bold $N$ value, is used in the reported offset results.
An italic $N$ value signifies which supercell is used for energy results.}
\end{table}

\begin{figure}
\includegraphics{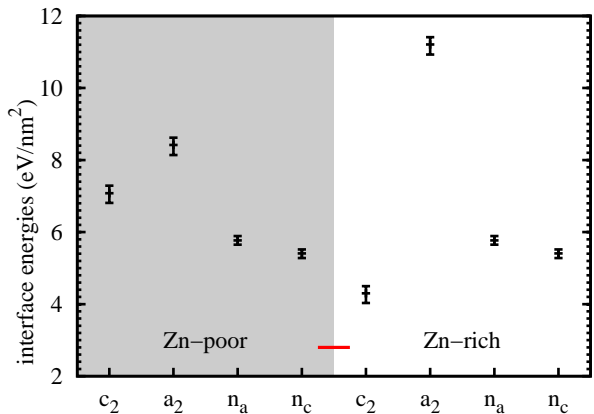}
\caption{\label{f:100energies}(Color online) Energies for (100) (black) and (110) (red) interfaces under Zn-poor (shaded) and Zn-rich (unshaded) growth conditions.
Error bars denote the minimum and maximum of four energy values calculated using simple combinations of supercell calculations.}
\end{figure}

Figure \ref{f:100energies} shows the means and the ranges of the calculated interface energies for the [100] direction, as well as the [110] result.

The most stable (100) interfaces are polar-compensated, with the exception (for the set of considered interfaces) of the abrupt cation termination $c_2$ under Zn-rich conditions ($E_{c2} = 4.3$ eV/nm$^2$). 
The calculations also show the $c_2$ interface to be more stable than the $a_2$ (Si-S) interface, for the full range of chemical potential space.
This is somewhat unexpected in light of  the measurements\cite{maierhofer91val,zhou97epi} discussed in Sec.~\ref{s:studies} which suggest that the (111) Si surface prefers a S adlayer to a Zn adlayer.
The fact that the formation of an interfacial SiS$_2$ layer is sensitive to growth environment (Sec.~\ref{s:studies}) indicates that the experimental growth conditions in these studies are likely to have been Zn-poor.
At these conditions, the $c_2$ interface energy is 7.1 eV/nm$^2$, still less than the $a_2$ energy of 8.4 eV/nm$^2$.
%
However, the calculation suggests a that defective interface such as $n_a$ or $c_n$ is the most stable under the experimental Zn-poor conditions.
This result is not in any serious conflict with experiments\cite{maierhofer91val,zhou97epi}.

\begin{figure}
\includegraphics{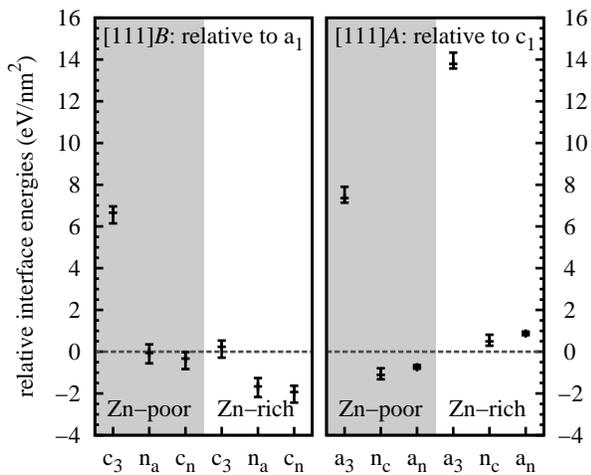}
\caption{\label{f:111energies}Left: Abrupt and polar-compensated $[111]B$ oriented interface energies relative to the $(111)B$ single-dangling-bond Si-S interface energy, $E_{a1}$. Right: Abrupt and polar-compensated $[111]A$ oriented interface energies relative to $E_{c1}$. Shaded regions show energies under Zn-poor growth conditions. Error bars show the extent of the minimum and maximum values for each data set.}
\end{figure}

\begin{figure}
\includegraphics{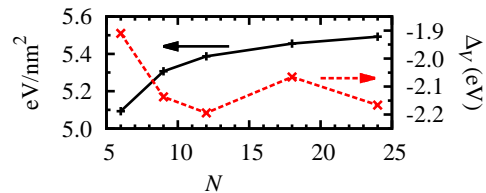}
\caption{\label{f:lengthconvergence}(Color online) Calculated (111) interface energy quantities $(E_{a1} + E_{c1})/2S$ (black solid) and VBO values (red dashed) for the $a_{1}$ interface, as a function of the number $N$ of quasi-unit cells in each slab.}
\end{figure}

Figure \ref{f:111energies} shows the calculated (111) interface energies relative to the $b=1$ abrupt interfaces for both $[111]A$ and $[111]B$ orientations.
Figure \ref{f:lengthconvergence} demonstrates convergent behavior for $(E_1 + E_2)/2$ in the $(a_{1}$,$c_{1})$ supercells with increasing slab size, $N$.
The difference between the $N=9$ and $N=24$ calculations is less than 0.2 eV/nm$^2$ despite these supercells having the highest field due to their strong
 donor-acceptor nature.
The calculation of convergence with large $L$ in this worst case is preferred to applying electrostatic energy corrections which either require surfaces (additional unknowns) or the assumption of transitivity (see Sec.~\ref{s:intro}).

The two considered reconstructions of the $a_{1}$ interface, $n_a$ and $c_n$, appear to be stable over nearly the entire range of chemical potential space ($3.7 \textrm{ eV}  < \mu_x < 5.4$ eV).
Additionally, the two reconstructions of the $c_{1}$ interface appear to be stable for over about half of the $\mu$-space.
Both the $a_{3}$ and $c_{3}$ interfaces are unstable when compared to the competing, polar-compensated interfaces, and are clearly unlikely to form.

Among the interfaces considered, we have the directional stability relations given by the energy inequalities:
\begin{align}
E_{[110]} < E_{[100]}, \qquad E_{[111]} < E_{[100]}, \label{e:directionalstability}
\end{align}
which hold separately for Zn-rich and Zn-poor growth conditions.
Appendix \ref{as:energycomparison} discusses how such energy inequalities must be determined.
While it is not possible to show that $E_{[110]} < E_{[111]}$, we may conclude that $E_{[110]} < (E_{[111]A} + E_{[111]B} )/2$, where all energies refer to the same growth conditions.

The calculated directional energy minima $E_{[100]}$ and $(E_{[111]A}+E_{[111]B} )/2$ are only upper bounds for experimentally realized minima, which almost certainly involve more complex defect structures and may be growth dependent due to metastable structures.
However, we may note that our directional stability relations for Si/ZnS differ somewhat from similar theoretical investigations of type IV/III-V interfaces:
studies\cite{lee90val,dandrea90sta} of Ge/GaAs, Si/GaAs, and Si/GaP interfaces based on short-period ($N = 3$ to 4) superlattices suggest that, for $N \geq 3$, $(E_{[111]A}+E_{[111]B} )/2$ is less than $E_{[110]}$.


\subsection{\label{s:bandoffsetresults}Band Offset Results}

\begin{figure}
\includegraphics{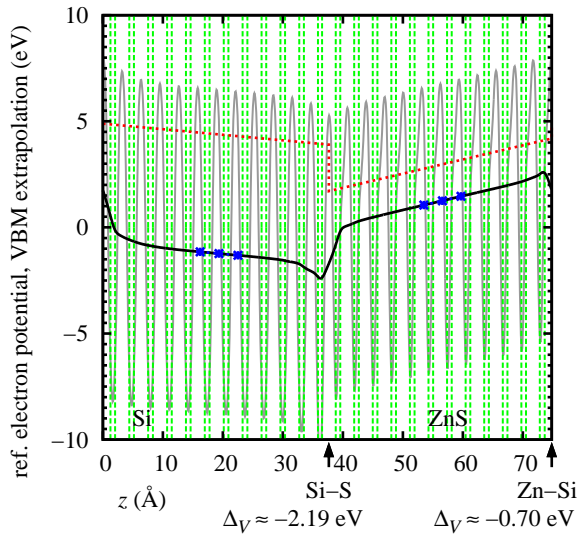}
\caption{\label{f:111potential}(Color online) Smoothed, plane-averaged electron potential (black) and extrapolated valence band maximum (VBM) (red dashed) for the $N=12$ [111] $(a_{1},c_{1})$ supercell. Ions and length are relaxed. Valence band slopes are calculated from potential averages (blue points) over the quasi-unit cell between like atomic planes near the slab centers (all atomic planes are shown, green dashed). The plane-averaged electron potential is shown in gray.
The width of the rapidly-varying potential region at the interface is greatly exaggerated by convolving the plane-averaged electron potential $V(z)$ with two unit-area top hat functions having lengths equal to the bulk quasi-unit cells; this smoothing is used to produce a smooth potential in both bulk regions.\cite{baroni89can}
Of all supercells for which $\Delta_V$ values are calculated, this supercell has the greatest value of $|\mathcal{E}_2 - \mathcal{E}_1|$, and therefore the greatest expected uncertainty in the extrapolation determining $\Delta_V$.}
\end{figure}

For the $N=12$ [111] $(a_{1},c_{1})$ supercell, Figure \ref{f:111potential} shows the plane-averaged, doubly convolved\cite{baroni89can} electrostatic reference potential and the parallel extrapolations lines used to estimate the $\Delta_V$ values.
The two polar interfaces form a donor-acceptor pair resulting in a large charge transfer.
The total charge in the Si-S interface region as determined by fields at the slab centers is $+0.0066$ $e$, while the $N=9$ version of the supercell has charge $0.0097$ $e$.
These are the largest charge values of all the supercells considered for the calculations of band offsets and interface energies, respectively.
Despite the uncertainty in the VBO calculation that is evident from the relatively large fields, the $\Delta_V$ results are within 0.1 eV of a similar calculation having an additional vacuum slab and $1/10$ the charge transfer (Table \ref{t:ifcdata}).
Figure \ref{f:lengthconvergence} shows relatively small variation of $\Delta_V$ as $N$ is varied.
The extrapolation technique demonstrated in Fig.~\ref{f:111potential} has been used to determine all valence band offset values.

\begin{figure}
\includegraphics{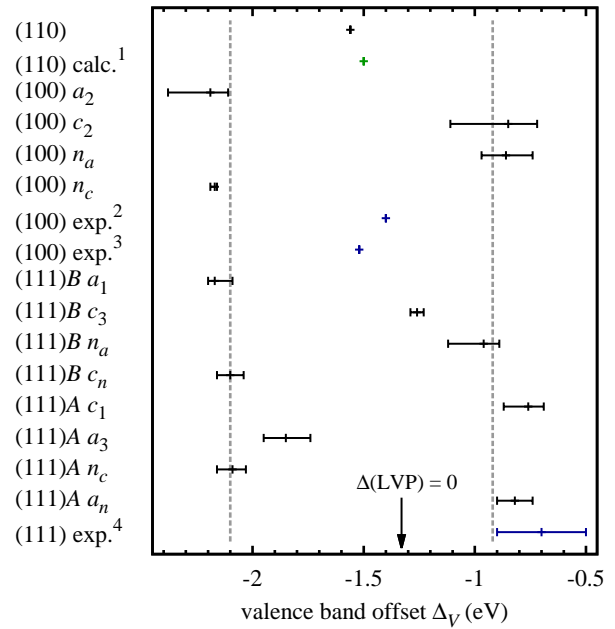}
\caption{\label{f:VBO}(Color online) Valence band offsets (from Si into ZnS). Values from other works are shown in green (calculation) and blue (experiment).
Error bars for our calculated values show the maximum and minimum values obtained from various supercell calculations.
The mean of the calculated (100) and (111) values less than -1.5 eV is shown by the dashed line at $-2.1$ eV.  The mean of the remaining calculated (100) and (111) values is show by the dashed line at $-0.92$ eV.
The arrow indicates the origin of the LVP offset, indicating that the interface dipole changes sign between the two peaks.
References are (1) Ref.~\onlinecite{li96ban}, (2) Ref.~\onlinecite{brar98ban}, (3) this value is mentioned via unpublished reference within Ref.~\onlinecite{lewyanvoon97ele}, and (4) Ref.~\onlinecite{maierhofer91val}.}
\end{figure}

Table \ref{t:ifcdata} shows $\Delta_V$, $\mathcal{E}$, and $\sigma$ data from individual supercell calculations.
Calculated and experimental $\Delta_V$ results are shown in Figure \ref{f:VBO}.
The (1$\times$1) [111] supercells used in the data sets have length $N=12$ rather than $N=9$.
The $a_{1}$ and $c_{1}$ data sets for $\Delta_V$ each include an $N=12$ supercell having an single interface and a vacuum slab with pseudo-hydrogen terminated surfaces.

The calculated $\Delta_V$ values vary from about $-0.8$ eV to $-2.2$ eV.
VBOs for both [100] and [111] directions fall into two groups of values, one including all abrupt anion terminations, centered at $-2.10$ eV, and the other including all abrupt cation terminations, centered at $-0.92$ eV.
Our calculated (110) $\Delta_V$ lies near the center of the entire distribution and agrees well with Li \emph{et al}.\cite{li96ban}
The experimental values\cite{brar98ban,lewyanvoon97ele} for the [100] direction also lies near the average of the two groups.
The simplest interface structures expected to have interface dipoles this small are 50/50 mixtures of $n_a$ and $n_c$ substitutions. 

In contrast to the other experimental results, the (111) VBO of Maierhofer \emph{et al}.\cite{maierhofer91val}~lies clearly in a single group (the abrupt cation group).
This indicates that the grown macroscopic (111) interfaces in these experiments had significant selectivity among the possible microscopic interfaces.

\begin{figure}
\includegraphics{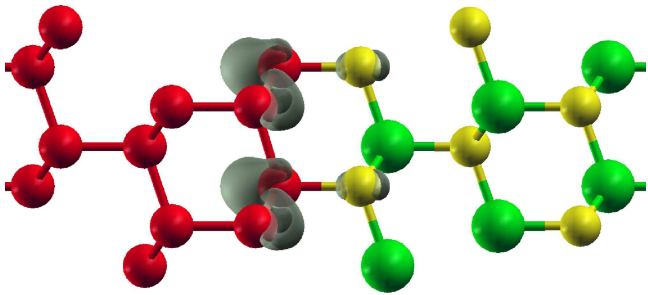}
\caption{\label{f:occsurfstate} (Color online) Charge density contour (transparent gray) of the occupied portion ($-0.43$ $e/S$) of the surface state at an $a_1$ interface in an $(a_1,c_1)$ supercell. The center of charge lies on the Si side of the interface region, and the state has visible traces of Si-S anti-bonding character. Atoms are Si (red), Zn (green), and S (yellow).}
\end{figure}

The bimodal behavior of the offsets calculated here can be grossly explained by a two charge model.
For anion terminated polar interfaces with electronic compensation, the extra electrons will go into a localized anti-bonding state formed primarily from Si and ZnS conduction bands.
The conduction band minimum (CBM) on the Si side is lower than the CBM on the ZnS side, and thus the electrons localize preferentially on the Si side of the interface.
(For the $a_1$ interface in simulations with and without ion relaxation, the center of mass of the occupied portion of the Si-S interface state (Figure \ref{f:occsurfstate}) can be verified by inspection of the $\psi_{n\mathbf{k}}$ to lie on the bulk side of the interfacial Si atomic plane.\cite{Note1})
A cation terminated polar interface has extra holes which also preferentially localize on the Si side of the interface, due to its higher VBM.
In the large $L$ limit, the free charge approaches the magnitude of the compositional charge, but with opposite sign, forming a dipole.
The location of the compositional charge is not well defined.
In particular, starting as in Harrison \emph{et al}.\cite{harrison78pol} with the insulating ``frozen bulk'' electronic state of Si on both sides of the interface (all Zn and S nuclei being created by transferring protons between adjacent Si atoms in one half plane), one can interpret the positive compositional charge at a Si-S interface as being generated by either the S or Si ion cores.
This symmetric situation can be modeled by placing the compositional charge at the center of the interface bond region.
Having compositional charge near the bond center and the free charge first moment located outside of the central bond region yields, for all abrupt polar interfaces, the correct sign of the dipoles and $\Delta$(LVP) values (Fig.~\ref{f:VBO}). 

Dipoles at ionically compensated interfaces can be analyzed in terms of changes to the electronically compensated abrupt interfaces.
The substituting nuclei can be constructed by moving the free charges into nuclei of the abrupt interface.
This construction changes electronic reconstruction into ionic reconstruction.
Within the model above, we may form a rule that the dipole changes sign if the free charge crosses the interface as it moves into the nuclei.
This rule can be seen to place each polar-compensating (100) and (111) interface shown in Fig.~\ref{f:VBO} onto the correct side of the distribution.

Alternatively, one may first determine the sign of the dipoles of the polar compensated interfaces using the arguments of Harrison \emph{et al}.\cite{harrison78pol}, and secondly determine the dipoles of the abrupt interfaces by moving charge out of the substituting nuclei.

The two charge model creates the type of dipole switching necessary to create the bimodal distribution.
However, the model ignores the important screening effect of the valence electrons on the dipoles.
With screening suppressed, the locations of the free charge component and the bond-centered compositional charge suggest abrupt interface dipole densities greater than 0.5 $e$\AA$/S$.
In contrast, the interface dipole densities corresponding to the $\Delta$(LVP) values of the left and right distribution peaks in Fig.~\ref{f:VBO} are  $p_{a} = 0.055$ $e$\AA~and $p_{c} = -0.029$ $e$\AA~per [111] surface unit, respectively.
We find that this reduction in $p$ by an order of magnitude persists at fixed ion interfaces, indicating the predominant screening comes from electrons in the interface region.

While the two charge model contains the dipole behavior necessary for a bimodal distribution,
there may not exist a single explanation for the relative narrowness of the two peaks.
The offsets for the triple bond interfaces $a_3$ and $c_3$ are clearly moved toward the center, reflecting that they have a different bonding geometry and atomic plane spacing  than the remaining single bond [111] interfaces. 
If common bonding geometry is a unifying factor, the very close agreement of the four double bond interfaces ([100] direction) with the six single bond interfaces may be somewhat coincidental.


The (110) VBO lies near the center of the distribution. This is expected as non-polar interface offsets are typically close to a mean or ``bulk'' offset value that exists independent of interface direction or details.
Highly non-uniform interfaces may be best handled with calculations in non-polar directions, or the use of direction-independent junction models such as that of of M\"onch\cite{monch07ele}.
In contrast to this ``bulk picture'' result is the large width (1.2 eV) of the entire offset distribution and the fact that grown interfaces\cite{maierhofer91val} can lie on a single peak.
The most transparent violations of the transitivity rule comes from the cation and anion terminated supercells, for which the opposing interface band offsets differ by 0.55 and 0.38 eV respectively (see Table \ref{t:ifcdata}). 
It is clear that the need to treat each interface individually will increase as high quality interfaces become more common.



Finally, it is a worthwhile exercise to consider the band offset results in light of the experiments by Maierhofer \emph{et al}.\cite{maierhofer91val}~for (111) interfaces.
Among the interfaces considered here there are three potentially stable (111) candidates having acceptable offsets: $n_a$, $a_n$, and $c_{1}$.
As noted in Sec.~\ref{s:studies}, Maierhofer \emph{et al}.\cite{maierhofer91val}~suggests that the first layer of ZnS in [111] growth is primarily $S$, lowering the likelihood of the $a_n$ and $c_{1}$ interfaces.
When ZnS is deposited on a cleaved (2$\times$1) (111) Si surface, as is done by Maierhofer \emph{et al}., the formation of an $n_a$ interface seems unlikely because it would involve either the presence extra Si atoms on the surface, or removing seven out of eight Si atoms in the first bilayer to achieve the effect of a $1/4$ Si substitution into ZnS.
It is perhaps most probable that the experiment created interfaces that are not  formed from the primitive substitutions considered here.
As mentioned previously however, the non-bulk dipole signifies a degree of selectivity within the universe of low energy microscopic interface structures.

\section{\label{s:conclusions}Conclusions}

Polar interfaces of traditional semiconductors contain the physical ingredients needed to produce the carrier and charge phenomena now commonly examined in complex oxide structures. 
Any compositional (polar) charge between slabs of sufficient thickness $L/2$ must be screened by induced bound charge, conventional polarization charge (e.g.~piezoelectric charge), and free charge. Induced bound charge and net charge go to zero as $L \rightarrow \infty$, while the remaining components generally reach non-zero asymptotes.
The most stable interfaces are almost always polar-compensated, containing no compositional charge and generally lacking the need for any free charge.
For this reason the predicted metallicity in semiconductor abrupt polar interfaces is avoided, as it is in most oxide interfaces.

The closely lattice matched Si/ZnS interfaces deserve investigation for both practical and theoretical reasons. The enthalpy studies here confirm that polar-compensated interfaces are typically more stable than abrupt interfaces.
In the polar [100] direction, only the abrupt Si-Zn interface under Zn-rich conditions has not been excluded by the set of primitive (single-substitution) polar-compensating defects considered here.
In the [111] direction, interfaces having three interface bonds per interface atom are clearly unstable under all conditions, and this is likely also true for interfaces with one interface bond per atom.


The band offset calculations performed here show that the abrupt and single-substitution polar-compensated (100) and (111) interfaces have a bimodal distribution of offsets, with peaks at VBO values of $-0.9$ eV and $-2.1$ eV.
The width of the distribution indicates that band offset transitivity is violated by over 1.2 eV.
The Si-Zn interfaces having (111)$A$ (single bond) and (111)$B$ (triple bond) orientations have a VBO difference of 0.5 eV, indicating that interface orientation plays nearly as large a role as the ionic character of an interface.
A measured (111) VBO value\cite{maierhofer91val} lies near the peak at $-0.9$ eV and shows a clear preference of interface dipole.
Measured (100) VBO values\cite{brar98ban,lewyanvoon97ele} correspond with the average value of the distribution, and indicate the presence of ordered or disordered polar compensating structures having no net dipole.
The existence of the bimodal interface dipole distribution is consistent with  a primitive two charge model considering the free charge and the compositional charge.

Finally we have shown that extrapolation of linearly varying potentials can produce band offsets that are consistent with supercell calculations involving negligible electric fields $\mathcal{E}$.
Furthermore, interface energy calculations calculated using cells with significant fields and interface charges are consistent with energy calculations using only low $\mathcal{E}$ supercells.
Nine or more atomic bilayers per slab provides adequate length for meaningful results.




\begin{acknowledgments}
This work has been supported by the National Science Foundation of the USA under Grant SOLAR DMS-1035513.
\end{acknowledgments}

\appendix
\section{\label{as:energycomparison}Directional Energies}

Here we note how energy inequalities such as Eq.~(\ref{e:directionalstability}) must be computed, assuming a given set of defective interfaces.
In the statement $E_{[111]} < E_{[100]}$, $E_{[100]}$ simply denotes the minimum computed (100) interface energy.
For the [111] direction however, algebraic considerations yield an uncertainty in $E_{-} \equiv (E_A - E_B)/2$ for interfaces $A$ and $B$ having respective orientations $[111]A$ and $[111]B$.
Since $E_{-}$ is unknown, we must assume that its actual value may hinder the inequality we wish to state.
We first determine the maximum of $\min(E_A = E_{+} + E_{-}, E_B = E_{+} - E_{-})$, $E_{+} \equiv (E_A + E_B)/2$ with respect to the $E_{-}$.
This maximum occurs at $E_{-} = 0$ and thus the candidate value for $E_{[111]}$ for this $A$ and $B$ is $E_{+}$.
Minimizing over interface pairs, the appropriate $E_{[111]}$ value is given in terms of calculable values as $[\min(\{E_{A}-E_{c1}\})+\min(\{E_{B}-E_{a1}\}) + (E_{c1}+E_{a1})]/2$.
For our set of interfaces, $E_{[111]}$ is $4.88$ and $4.68$ eV/nm$^2$ for Zn-poor and Zn-rich conditions, respectively.
One cannot establish $E_{[111]} > E_{D}$ for any direction $D$ because the appropriate $E_{[111]}$ for this inequality involves a minimum with respect to $E_{-}$, rather than a maximum; here $E_{[111]}$ must be allowed to go to its lowest possible value, zero (excluding negative interface energies).
It follows that for two general directions, no directional energy inequality may be written without using an alternate means of interface energy computation (see Sec.~\ref{s:intro}).


\bibliography{paper}

\end{document}